\documentclass[aps,prl,floatfix,10pt,amssymb,amsfont,amsmath,superscriptaddress,float,nofootinbib,longbibliography,nobalancelastpage,twocolumn]{revtex4-2}
\usepackage{graphicx}
\usepackage{dcolumn}
\usepackage{bm}
\usepackage{braket}
\usepackage{color}
\usepackage[usenames,dvipsnames,svgnames,table,x11names]{xcolor}
\usepackage[normalem]{ulem} 

\newcommand{\Tr}{{\mathop{\mathrm{Tr}}}}
\usepackage{hyperref}
\begin{document}

\preprint{APS/123-QED}

\title{Fast extraction of universal data from temporal spectrum:\\
Applications to Haagerup CFT and monopole scattering}
\author{Tokiro Numasawa}
\affiliation{Institute for Solid State Physics, The University of Tokyo, Chiba, 277-8581, Japan}
\author{Boris De Vos}
\affiliation{Department of Physics and Astronomy, Ghent University, Krijgslaan 281, 9000 Gent, Belgium}
\author{Masataka Watanabe}
\affiliation{Faculty of Science, The University of Tokyo, Tokyo 113-0033, Japan}

\author{Atsushi Ueda}
\affiliation{Department of Physics and Astronomy, Ghent University, Krijgslaan 281, 9000 Gent, Belgium}


\date{\today}

\begin{abstract}
Extracting precise universal data on conformal field theory (CFT) on the lattice traditionally requires large-scale simulations of the ground states and excitations of critical systems. Here, we propose an alternative accurate and fast algorithm, where we read off the CFT spectrum from the transfer matrix spectrum of an infinite matrix product state after a short imaginary-time evolution.  We apply this method to extract the charge-resolved spectrum of a Haagerup CFT to higher levels and obtain a central charge $c = 2.137(27)$. Moreover, we show that the Maldacena--Ludwig monopole boundary can be realized as the ground state of the Fidkowski--Kitaev Hamiltonian after space-time rotation.
\end{abstract}

\maketitle
\paragraph{Introduction.}
“More is different.” In his celebrated 1972 essay, Anderson challenged the reductionist view that understanding the fundamental laws of nature is sufficient to explain all physical phenomena~\cite{Anderson:1972pca}. While breaking a system into its elementary constituents may reveal its microscopic laws, collective phenomena can exhibit entirely new organizing principles that are invisible at the microscopic level. Remarkably, in quantum many-body physics, more can also mean simpler. At criticality, systems with vastly different microscopic realizations exhibit the same universal behavior, characterized by only a handful of parameters. Identifying these universal quantities is therefore a central goal in understanding quantum many-body systems. In one dimension, conformal invariance provides a particularly powerful manifestation of universality: the low-energy spectrum becomes universal up to scaling~\cite{Cardy1984,BloteCardyNightingale1986}. Moreover, $(1+1)$-dimensional {\it boundary} conformal field theory is ubiquitous because it can also describe the low-energy physics of impurity problems in {\it higher dimensions}, where an $s$-wave reduction yields an effective $(1+1)$-dimensional theory on the radial half-line.

Ironically, critical systems are among the most challenging for numerical simulations. The divergence of correlation lengths and the associated growth of entanglement make even ground-state calculations computationally demanding for conventional density matrix renormalization group (DMRG) methods~\cite{PhysRevLett.69.2863,Tagliacozzo2008}. Extracting the excitation spectrum is considerably more difficult, typically requiring additional variational calculations targeting excited states~\cite{PhysRevB.73.014410,qpansatz2,qpansatz}. Although alternative approaches can directly access excitation spectra~\cite{Chepiga_2017}, resolving the complete low-energy spectrum across different symmetry sectors remains challenging. These difficulties become particularly severe for theories with larger central charges, where the increased entanglement requires substantially greater computational resources~\cite{Pollmann2009}.

In this Letter, we introduce a fast and accurate method for extracting the low-lying universal spectrum of critical quantum systems using the temporal spectrum, previously explored in the context of real-time evolution~\cite{luca1,luca2,luca3}. Related temporal entanglement spectra have also recently been investigated in Euclidean-evolved critical spin chains~\cite{YangHuang2026}.
BCFT spectra have also been extracted using the boundary
tensor renormalization group~\cite{Iino:2019vxd,Iino:2019rxt}.
Here, we demonstrate that the universal conformal spectrum can be obtained directly from the transfer matrix of a state evolved in imaginary time~\cite{Vidal2004,PhysRevLett.107.070601}, without explicitly constructing excited states. This approach is particularly advantageous for critical theories with larger central charges, where conventional variational methods require large bond dimensions and costly iterative optimization. We demonstrate its power by investigating two challenging examples: the Haagerup CFT with $c\approx2$~\cite{Vanhove2022,Huang2022,Liu_2023} and the Maldacena--Ludwig wall with $c=4$~\cite{Maldacena_1997,Affleck:1993np,Tachikawa:2026cxd}.

Our method also provides an efficient way to construct boundaries and see their spectrum, which involves chiral symmetries originating from the lattice translation symmetries \cite{PhysRevB.36.5291,Furuya:2015coa,Cheng:2022sgb,Seiberg:2023cdc}.
A translation-invariant spatial boundary is tricky because the boundary itself breaks translation symmetry, but we can impose the exact translation invariance for boundary states on a lattice.
In the $c\approx 2$ Haagerup CFT, we study the boundary states that break this symmetry, and in the $c=4$ theory we study the Maldacena--Ludwig boundary states, which are characterized as the Majorana translation-invariant boundary.
We realize the Maldacena–Ludwig boundary universality class using an imaginary-time-evolved Fidkowski–Kitaev product state and verify its identical- and mixed-boundary spectra.

\paragraph{Open-closed duality on a lattice.}
\begin{figure}[tb]
    \centering
    \includegraphics[width=\linewidth]{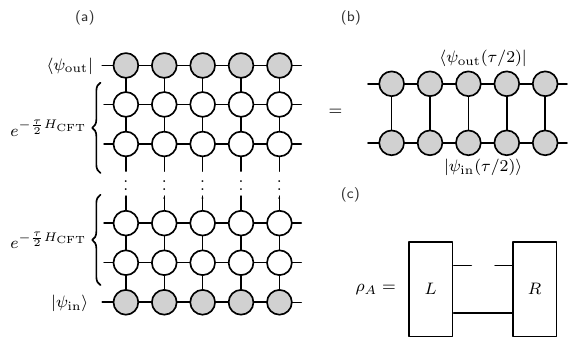}
    \caption{(a) Tensor-network representation of the Euclidean path integral. (b) The path integral becomes an overlap of the two time-evolved states. (c) The left and right leading eigenvectors correspond to the ground state of the rotated Hamiltonian. The reduced density matrix can be constructed by sewing one leg.}
    \label{fig:pathintegral}
\end{figure}

    
Open-closed duality is an important consistency condition of boundary conformal field theories (BCFTs).
It comes from the fact that the Euclidean time and space can be treated equally in Euclidean path integrals.
The thermal partition function of the CFT on an interval of length $L$ and inverse temperature $\beta$, with boundary conditions $a$ and $b$ at either end can be interpreted in two different ways:
\begin{align}
    \Tr(e^{-\beta H_{ab}^{\rm open}(L)}) = \braket{b|e^{-\frac{L}{v}H^{\rm closed}(v\beta)}|a},
    \label{eq:open-closed}
\end{align}
where $H_{ab}^{\rm open}(L)$ denotes the Hamiltonian on the interval of length $L$, $H^{\rm closed}(v\beta)$ the Hamiltonian on a circle of circumference $v\beta$, and $\ket{a}$, $\ket{b}$, corresponding boundary states.
We write $\tau \equiv L/v$ for the Euclidean time over which the boundary states propagate in the closed channel.
Note that we have retained the speed of light, $v$, for later reference.
Furthermore, it is customary to define dimensionless conformal weights by conformally mapping the former to a half-annulus and the latter to a full annulus
\begin{align}
    \Tr\left[\left(e^{-\frac{\pi v\beta}{L}}\right)^{\Delta_{\rm open}-\frac{c}{24}}\right] = \braket{b|\left(e^{-\frac{4\pi L}{v\beta}}\right)^{\frac{1}{2}(\Delta_{\rm closed}-\frac{c}{12})}|a},
    \label{eq:open-closed-2}
\end{align}
where $\Delta_{\rm open, closed}$ are called conformal weights, while $c$ is the central charge.
Put differently, we strip the dimension off by scaling $L/v$ to $\pi$ in the open channel, or $\beta$ to $2\pi$ in the closed channel.

Our proposal is to use the open-closed duality to extract physical data of BCFTs, exploiting the fact that numerically estimating the RHS of \eqref{eq:open-closed} is much easier than the LHS.
This is because boundary states are easy to implement as translation-invariant, short-range entangled states \cite{Miyaji:2014mca,Cardy:2017ufe,Cho:2016xjw,Numasawa:2018grg} and as such should be written as an infinite matrix product state (iMPS),
\begin{align}
    \begin{gathered}
        \ket{a} = \sum_{i_1,\dots, i_N} \Tr (A_{1}^{i_1}  \cdots A_N^{i_N}) \ket{i_1, \dots , i_N}\\
        \ket{b} = \sum_{i_1,\dots, i_N} \Tr (B_{1}^{i_1}  \cdots B_N^{i_N}) \ket{i_1, \dots , i_N}
    \end{gathered}
\end{align}
where $A^{i}_j= A^{i}_{j+s}$ and $B^{i}_j= B^{i}_{j+s}$.
We identify $\beta \equiv \epsilon N/v$, where $\epsilon$ is the lattice constant.
Both boundary states can be further evolved by using the spin-chain Hamiltonian $H$ (with some overall normalization characterized by a linear slope of the dispersion relation at low-energy given by velocity $v$) using the infinite time-dependent variational principle (iTDVP)~\cite{PhysRevLett.107.070601} as
\begin{align}
    \begin{gathered}
        e^{-\tau H/2}\ket{a} = \sum_{i_1,\dots, i_N} \Tr (\tilde{A}_{1}^{i_1}  \cdots \tilde{A}_N^{i_N}) \ket{i_1, \dots, i_N}\\
        e^{-\tau H/2}\ket{b} = \sum_{i_1,\dots, i_N} \Tr (\tilde{B}_{1}^{i_1}  \cdots \tilde{B}_N^{i_N}) \ket{i_1, \dots, i_N}
    \end{gathered},
\end{align}
where $\tilde{A}$ satisfies periodicity in its subscript, as iMPS should.
The RHS of \eqref{eq:open-closed} is therefore given as
\begin{align}
    \begin{split}
            &\braket{b|e^{-\frac{L}{v}H^{\rm closed}(v\beta)}|a}\\
        &\qquad\qquad=\lim_{N\to \infty}\sum_{i_1,\dots, i_N}\Tr \left(\prod_{k=1}^N(\tilde{A}^{i_k}_k\otimes (\tilde{B}^{i_k}_k)^*)\right).
    \end{split}
\end{align}
We can therefore conclude that
\begin{align}
    H^{\rm open} = -\frac{v}{s\epsilon}\log \left[\prod_{k=1}^s(\sum_{i_k}\tilde{A}^{i_k}_k\otimes (\tilde{B}^{i_k}_k)^*)\right],
\end{align}
which can be used to derive conformal dimensions of operators in BCFTs. In practice, we can find the velocity $v$ by identifying the energy level of the stress-energy tensor.

\begin{figure}[tb]
    \centering
    \includegraphics[width=\linewidth]{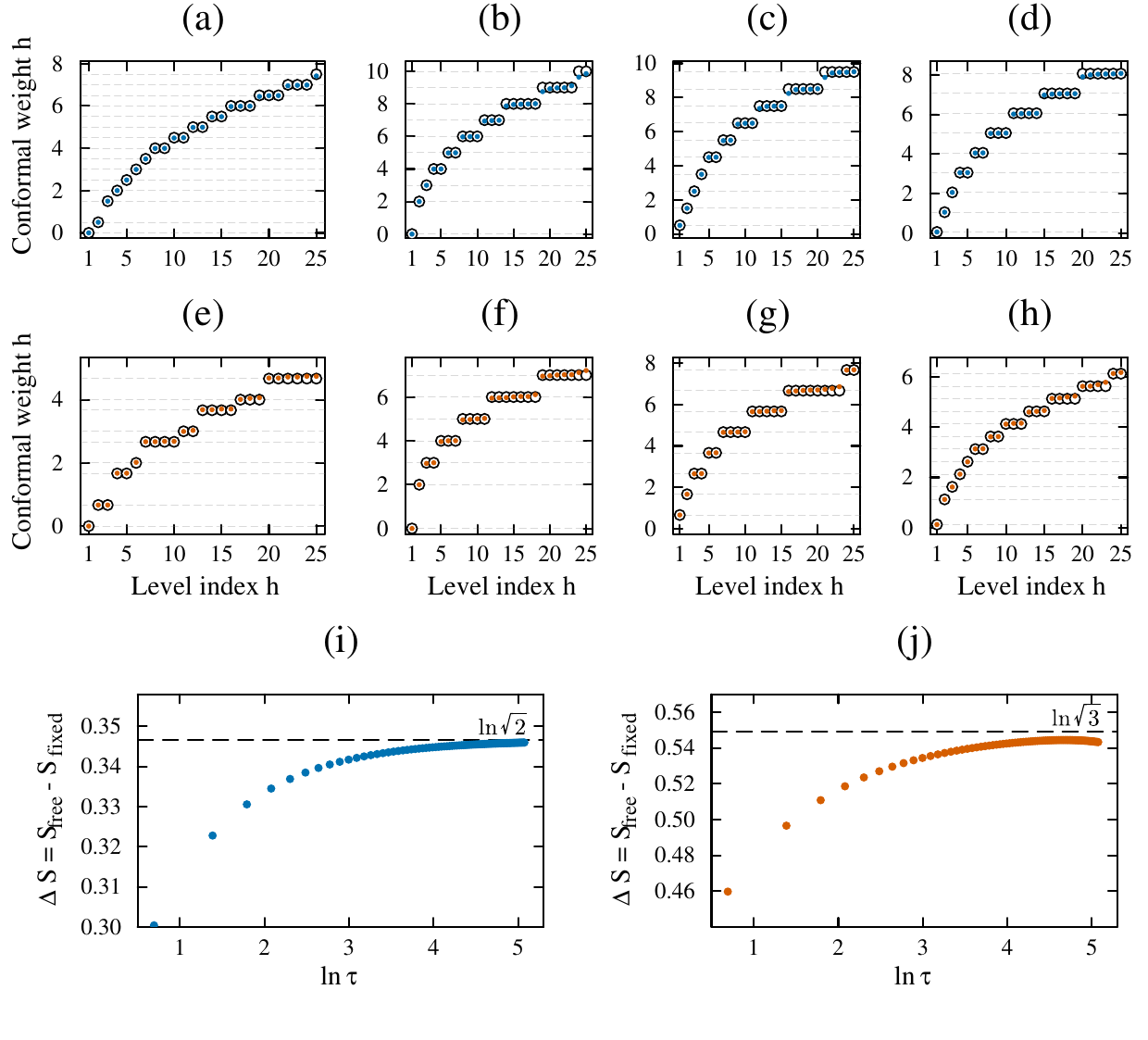}
    \caption{Conformal weights extracted at $\tau=12$ for the critical Ising chain with (a) free-free, (b) up-up, (c) up-down, and (d) free-up boundary conditions ($\chi=160$), and for the critical three-state Potts chain with (e) free-free, (f) A-A, (g) A-B, and (h) free-A boundary conditions ($\chi=210$). Open black circles indicate the conformal weights predicted by BCFT, while colored dots show the numerical results. (i, j) The AL boundary entropy difference of the free-boundary condition for the Ising and Potts model. }
    \label{fig:spectrum_benchmark}
\end{figure}

\paragraph{Benchmark results.}
We begin by demonstrating the power of the method in the paradigmatic critical spin chain, the transverse-field Ising model. In the low-energy limit, this theory admits three conformal boundary conditions: up, down, and free. We prepare these boundary states as ground states in the presence of longitudinal or transverse fields. Using them as the initial and final states of the imaginary-time evolution, we extract the BCFT spectra for various choices of boundary conditions. The results are shown in Fig.~\ref{fig:spectrum_benchmark}(a–d), where the blue dots denote the numerically obtained conformal weights and the open black circles the exact BCFT predictions. Even at modest numerical cost, the agreement is excellent up to level 6.
We can carry out the same analysis for the critical three-state Potts model. In this case, the fixed boundary condition has three possible orientations, $A$, $B$, and $C$. As shown in Fig.~\ref{fig:spectrum_benchmark}(e–h), the method again reproduces the expected BCFT spectra with high accuracy. Furthermore, the numerically extracted Affleck--Ludwig (AL) boundary entropies  \cite{Affleck:1991tk} also agree well with the analytical predictions, as shown in Fig.~\ref{fig:spectrum_benchmark}(i,j).

\paragraph{Application to a Haagerup anyon chain.} 
The free and fixed boundary states above can be written as short-range entangled states, in particular product states or more generally ground states of a gapped Hamiltonian.
Such a state is a conformal boundary state smeared over an imaginary time and dressed by boundary perturbations \cite{Cardy:2017ufe}.
The imaginary time evolution preserves the symmetries this state and the Hamiltonian have in common, so the resulting boundary condition is stable against the perturbations sharing these symmetries.

We now wish to apply this method to a $\mathcal{H}_3$ Haagerup critical model \cite{Vanhove2022, Huang2022, Liu_2023}.
This fusion category has simple objects $\{1, \alpha, \alpha^2, \rho, \alpha\rho, \alpha^2\rho\}$.
The Hamiltonian is
\begin{equation}
    H = -\sum_i P^\rho_{i, i+1},
\end{equation}
where we project onto the $\rho$ channel of $\rho \otimes \rho = 1 \oplus \rho \oplus \alpha\rho \oplus \alpha^2\rho$, whose quantum dimension is $d_\rho = \tfrac12(3+\sqrt{13})$.
This chain is critical with $c \approx 2$, yet its CFT is unidentified.
From the fusion rules, it is immediately clear that we cannot start from a product state.
The closest analogue which maximizes the number of vacuum bonds for $\mathcal{H}_3$ is the trimer with bonds $(\rho, \rho, 1)$, since it has been found that translation acts at low energy as an emergent $\mathbb{Z}_3$ symmetry.
We call this candidate smeared boundary state $\ket{\rm A}$.
A one-site translation defines $\ket{\rm TA}$, and translating once more $\ket{\rm T^2A}$~\footnote{A priori, there are not necessarily distinct boundary conditions since we do not know the nature of this $\mathbb{Z}_3$ symmetry in the CFT.}.

\begin{figure}[tp]
    \centering
    \includegraphics[width=\columnwidth]{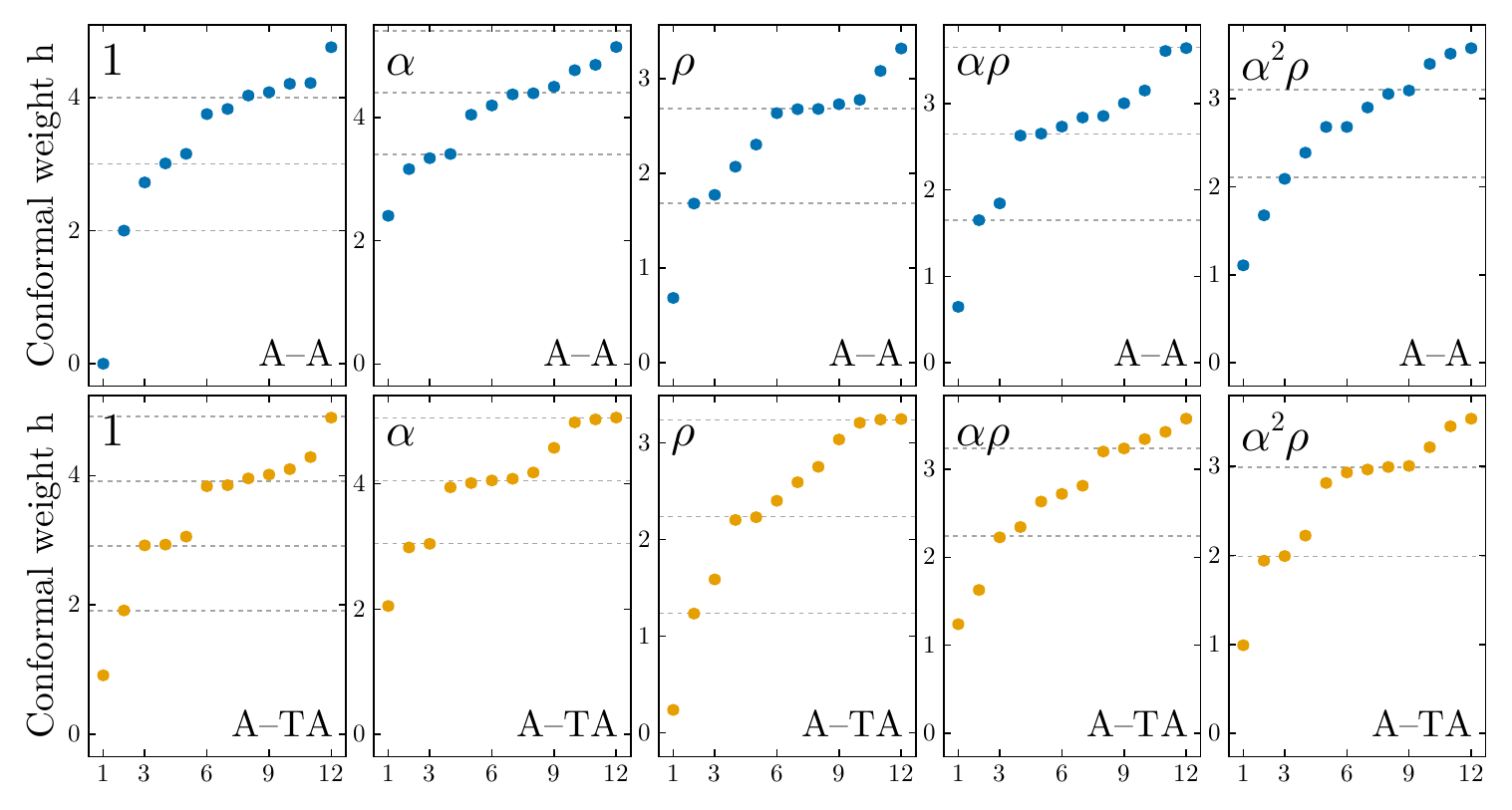}
    \caption{Extracted conformal weights of the considered $\mathcal{H}_3$ Haagerup CFT for two sets of boundary conditions, one where they are equal (A--A, top row), and one where they are not (A--TA, bottom row), at $\tau = 32$. The black dotted lines denote the descendants that are integer-spaced from the lowest primary within the tower. The $\alpha^2$ spectra match the $\alpha$ exactly in the A--A case, and up to $2 \times 10^{-3}$ for A--TA.}
    \label{fig:haagerup_spectrum}
\end{figure}

We show in Fig.~\ref{fig:haagerup_spectrum} the BCFT spectrum for boundary conditions A--A and A--TA at $\tau = 32$.
The appendix tabulates these values.
We normalize the spectrum by setting the conformal weight of the stress tensor to 2 in the identity tower.
We learn from these spectra that $\ket{\rm A}$ is a simple boundary condition since there is exactly one $h = 0$ level across all symmetry sectors.
Additionally, $\ket{\rm A}$ is stable against local perturbations, since the relevant perturbations are all charged in the non-invertible part of $\mathcal{H}_3$.
Another interesting observation is the $h = 0.990$ primary in the $\alpha^2\rho$ A--TA spectrum, which is a potential candidate for a spin-1 current in the twisted sector~\cite{Liu_2023}.

We can extract the velocity and central charge from the transfer spectrum $\lambda_n$ and energy density $e$ evolved over an imaginary time $\tau$, which corresponds to an interval of length $L = v(\tau+\tau_0)$, with $\tau_0$ an extrapolation length, and an infinite closed circle ($\beta\to\infty$), through

\begin{align}
    \begin{aligned}
   -\frac{v}{L_x}\ln |\lambda_n| &= E^{\rm open}_{ab, n} - E^{\rm open}_{aa, 0} = \frac{\pi v}{L}\Delta_{\mathrm{open},n} \, , \\
    e &= \frac{\langle H^{\rm closed}\rangle}{v\beta} = e_\infty + \frac{\pi v c}{24L^2}, \label{eq:TransferEigen}
    \end{aligned} 
\end{align}
where $L_x = 3$ is the length of the unit cell, $\Delta_{\rm open}$ is measured from the $aa$
ground state, and $e$ is taken for $a = b$ \cite{Cardy:1984bb, Cardy:1989ir, Calabrese_2007}.
We find $v = 1.016(5)$ and $c = 2.137(27)$, which is in line with previous results \cite{Vanhove2022, Huang2022, Liu_2023}.

The spectra in Fig.~\ref{fig:haagerup_spectrum} are the two unique sets; the other equal-boundary-condition cases give the same spectrum, and the A--T$^2$A spectrum matches that of A--TA but with $\alpha \leftrightarrow \alpha^2$.
These data have consequences for the candidate CFT behind this critical model.
We can exclude a tensor product of CFTs since the stress tensor is multiplicity-free.
Similarly, WZW current algebras are excluded due to the absence of $h = 1$ local currents in the vacuum sector \cite{Recknagel:1998ih}.
For the same reason, we can reject the W$_3$ (W$_4$) algebra due to just one (two) level(s) for $h = 3$ (4) in the identity tower.
In principle, cosets which project out currents up to spin-4 are allowed, or the CFT can be irrational.

\paragraph{Maldacena--Ludwig monopole scattering.}
In this section, we construct the Maldacena--Ludwig (ML) boundary states \cite{Maldacena_1997} on a lattice and apply our method to study the BCFT spectrum.
We consider eight copies of critical Majorana chains\cite{Kitaev:2000nmw}, whose continuum limit is a theory of eight massless Majorana fermions with central charge $c=4$. 
At each lattice site $j$, we combine the eight Majorana operators into four complex fermions,
$
f_{j,\alpha}
=\frac{1}{2}(\chi_j^{2\alpha-1}+i\chi_j^{2\alpha}),
 \alpha=1,\dots,4,
$
with 
$\{\chi_j^a , \chi_{k}^b \} = 2\delta^{ab}\delta_{jk}$. 
The states $\ket{n_1n_2n_3n_4}$ below denote the corresponding occupation-number basis.
\begin{align}
\ket{\psi_{\text{FK}}}  = \frac{1}{\sqrt{2}} (\ket{0000} - \ket{1111}), \ \ket{\Phi_{\text{FK}}} = \otimes _{i = 1}^L \ket{\psi_{\text{FK}}}_i. 
\end{align}
This is the ground state of the Fidkowski--Kitaev (FK) Hamiltonian \cite{Fidkowski:2009dba,You:2014vea}
\begin{align}
H = H_{8\text{Kitaev}} + g H_{\text{FK}},
\end{align}
at $g \to \infty$.
Then, we evolve the state by the $g=0$ critical Kitaev chain Hamiltonian:
\begin{align}
\ket{\Phi_{\text{FK}}(\tau)} = e^{- \frac{\tau}{2} H_{8\text{Kitaev}}}\ket{\Phi_{\text{FK}}}. 
\end{align}
Although $\ket{\Phi_{\mathrm{FK}}}$is a product state across lattice sites, its local constituent $\ket{\psi_{\mathrm{FK}}}$ is not a free-theory ground state. 

To motivate the identification of the FK preparation with the ML boundary condition, we compare their symmetry properties. The ML boundary condition preserves the chiral fermion parity $(-1)^{F_L}$, whose anomaly cancels for eight Majorana fermions. At the $\mathrm{Spin}(7)$-symmetric point considered here, it also preserves the subgroup associated with the fixed-point subalgebra of the order-two triality automorphism exchanging $\bm{8}_v \leftrightarrow \bm{8}_s$ representations. 
 
The FK product state preserves $\mathrm{Spin}(7)$, one-site Majorana translation, and CRT \footnote{Combining the translation and the $\mathrm{CRT}$ symmetry, the $N$ mod $8$ anomaly of $(-1)^{F_L}$ is matched in the UV and IR \cite{Seiberg:2025zqx}. }. 
In the scaling limit, one-site Majorana translation acts as a spatial translation combined with chiral fermion parity, which we denote by $(-1)^{F_L}$.
The FK preparation thus preserves microscopic counterparts of the symmetries of the ML boundary condition. 
This correspondence motivates identifying its infrared boundary condition with the ML boundary condition.

In the continuum CFT, the BCFT spectrum with the Maldacena--Ludwig boundary is \cite{Maldacena_1997,Ishikawa:2005ea,Fukusumi:2021zme}
\begin{align}
&Z_{\mathrm{ML},\mathrm{ML}}(q) \notag \\
&=\chi_0^{SO(8)_1} + \chi_{\bm{8}_v}^{SO(8)_1}  + \chi_{\bm{8}_s}^{SO(8)_1}  + \chi_{\bm{8}_c}^{SO(8)_1}  .
\end{align}
where $\chi^{SO(N)_1}_\rho(q)$ is a character for the current algebra of $\mathfrak{so}(N)_1$.
The ML boundary condition indeed preserves a symmetry larger than $\mathrm{Spin}(7)$ that is given by the twisted diagonal symmetry $SO(8)_\omega \subset SO(8)_L \times SO(8)_R$, with generators embedded as $(J_L^a,\omega(J_R^a))$ \cite{Fukusumi:2021zme,Smith:2020nuf}. Here $a$ is an adjoint index, and $\omega$ is the order-two outer automorphism of $\mathfrak{so}(8)$ that exchanges the vector representation with one of the spinor representations. In our lattice realization, only the $\mathrm{Spin}(7)$ subgroup associated with the fixed-point subalgebra of $\omega$ is preserved; its generators satisfy $\omega(J^a)=J^a$.

To study the mixed ML--free boundary problem, we introduce a fully dimerized state either in the trivial phase or SPT phase in each layer as an initial state $\ket{\Phi_{\mathrm{free}}}$
corresponding to the free conformal boundary condition
\begin{align}
\ket{\Phi_{\mathrm{free}}(\tau)} =
e^{-\frac{\tau}{2}  H_{8\mathrm{Kitaev}}}
\ket{\Phi_{\mathrm{free}}}.
\end{align}
The ML–ML spectrum is obtained from the transfer matrix of $\ket{\Phi_{\mathrm{FK}}(\tau)}$, whereas the ML–free spectrum is obtained from the mixed transfer matrix formed from $\ket{\Phi_{\mathrm{FK}}(\tau)}$ and $ \ket{\Phi_{\mathrm{free}}(\tau)}$, following the construction above.

\begin{figure}[tb]
    \centering
    \includegraphics[width=1.0\columnwidth]{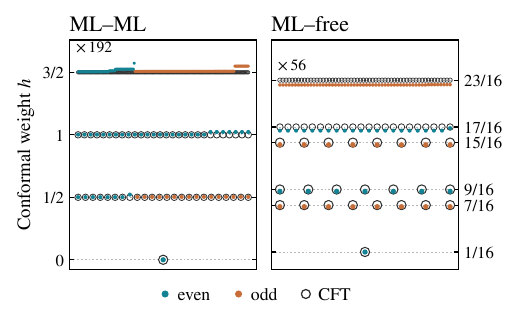}%
    \caption{Low-lying conformal weights extracted from the MPS transfer matrix for ML--ML (left) and ML--free (right) boundary conditions at imaginary time $\tau=4$.
    The velocity is $v=2$ for free Kitaev chains.
    Blue and orange dots denote numerical levels in the even and odd fermion-parity sectors, respectively; open circles denote the BCFT predictions. States within each multiplet are spread horizontally to show their multiplicity.
    } 
    \label{fig:MLspectrum}
\end{figure}
Free boundary conditions preserve the diagonal $SO(8)_V \subset SO(8)_L \times SO(8)_R$ symmetry with generators embedded as $(J_L^a,J_R^a)$.
The entire system is now invariant only under $SO(8)_{\omega} \cap SO(8)_V = \mathrm{Spin}(7)$ subgroup associated with the fixed-point subalgebra of $\omega$.
It is consequently useful to express the mixed-boundary partition function in terms of the $\mathrm{Spin}(7)_1$ and Ising characters associated with the conformal embedding
$SO(8)_1\supset \mathrm{Spin}(7)_1\times\mathrm{Ising}$ \cite{Affleck_1995,Affleck:1991yq}.
Using this embedding, the partition function with mixed boundaries is
\begin{align}
&Z_{\mathrm{ML},\mathrm{free}}(q) \notag \\
&=\big[\chi_0^{\text{SO(7)}_1}  + \chi_{\bm{7}}^{\text{SO(7)}_1} \big]\chi_\sigma^{\text{Ising}} +  \chi_{\bm{8}}^{\text{SO(7)}_1}  \big[\chi_0^{\text{Ising}}  + \chi_\epsilon^{\text{Ising}} \big].
\end{align}
Figure~\ref{fig:MLspectrum} compares the low-lying transfer-matrix spectra with the BCFT predictions for the ML–ML and ML–free boundary conditions. For ML–ML, the first excited states cluster near $h=1/2$: eight are even under fermion parity and sixteen are odd. Their total multiplicity of 24 agrees with the contribution of the three eight-dimensional primaries $\bm{8}_v,\bm{8}_s ,\bm{8}_c $ to $Z_{\mathrm{ML},\mathrm{ML}}$.

We use the numerically determined ML–ML ground-state level as the common reference for both spectra as we did in Eq.\eqref{eq:TransferEigen}. With this subtraction, the lowest ML–free level lies near $h=1/16$, providing an independent numerical check of the predicted conformal weight of the lowest boundary-condition-changing operator. This level is nondegenerate and even under fermion parity. The next two levels contain eight odd states near $h=7/16$ and seven even states near $h=9/16$, in agreement with $Z_{\mathrm{ML},\mathrm{free}}$. Together, the two spectra support the identification of the infrared boundary condition realized by the FK product state with the Maldacena–Ludwig boundary condition.

\paragraph{Conclusion and outlook. }
In this Letter, we have introduced an efficient method for extracting universal CFT data from the thermal partition function. By exploiting open–closed duality, we obtain the BCFT spectrum from the transfer matrix of an imaginary-time-evolved state, with boundary conditions encoded by translationally invariant, short-range-entangled infinite MPSs. Our method accurately resolves conformal spectra up to relatively high levels at modest computational cost. It is particularly powerful for critical theories with larger central charges, as demonstrated by our applications to the Haagerup CFT with $c\approx 2.1$ and the Maldacena–Ludwig monopole setup with $c=4$. These results establish a promising route to exploring challenging critical theories that remain computationally demanding for conventional DMRG-based approaches.

The applicability of the rotated framework extends beyond critical systems. For instance, a $90^\circ$ rotation of the real-time evolution of the transverse-field Ising model generally yields a non-unitary transfer matrix with complex couplings~\footnote{At criticality, the rotated Ising model has couplings $\tilde{J} = \tilde{g} = \frac{\pi}{4} - \frac{i}{2}\ln[\tan(Jt)]$, as derived in the Supplemental Material.}. This non-unitarity can effectively suppress certain modes, potentially explaining why contracting the quantum transfer matrix can offer improved numerical accuracy compared with direct real-time evolution. More broadly, our results suggest that spacetime rotation provides not only a new perspective on quantum dynamics but also a practical computational tool for studying strongly correlated quantum systems.
\paragraph{Acknowledgements.}
We have benefited from discussions with Hosho Katsura, Yuji Tachikawa, Shinsei Ryu, Shuhei Ohyama, Shu-Heng Shao, Zixia Wei, and Tadashi Takayanagi.
B.D. was supported by BOF-GOA (Grant No.\ BOF23/GOA/021).
A.U. was supported by FWO Junior Postdoctoral Fellowship (grant No. 12AHS26N). 
T.N. is supported by MEXT KAKENHI Grant No.~23K13094 and 24H00944 and by JST PRESTO Grant No.~JPMJPR2359.
M.W. is supported
by a Grant-in-Aid for JSPS Fellows No. 22KJ1777 and by a Grant-in-Aid for Early-Career
Scientists No. 25K17387.
\bibliography{apssamp}

\clearpage
\onecolumngrid
\begin{center}
\textbf{Supplemental Material}
\end{center}
\vspace{0.2cm}
\twocolumngrid
\section{Details on the Ising example} 
Here we describe the Ising model example to demonstrate how our lattice open-closed duality works.

In the continuum Ising CFT, we have three conformal boundary conditions labeled by $a =+, -,f$ \cite{Cardy:1989ir, Lewellen:1991tb}.
$+,-$ represents the fixed boundary condition whereas $f$ is the free boundary condition as will become explicit below in the lattice counterpart.
Correspondingly, we have three Cardy states
\begin{align}
\ket{B_+} &= \frac{1}{\sqrt{2}} \ket{I}\rangle + \frac{1}{\sqrt{2}} \ket{\epsilon}\rangle + \frac{1}{2^{\frac{1}{4}}} \ket{\sigma}\rangle  \notag \\
\ket{B_-} &= \frac{1}{\sqrt{2}} \ket{I}\rangle + \frac{1}{\sqrt{2}} \ket{\epsilon}\rangle - \frac{1}{2^{\frac{1}{4}}} \ket{\sigma}\rangle  \notag \\
\ket{B_f} &=  \ket{I}\rangle -  \ket{\epsilon}\rangle 
\end{align}
Here $\ket{I}\rangle, \ket{\epsilon}\rangle, \ket{\sigma}\rangle$  are Ishibashi states \cite{Ishibashi:1988kg,Onogi:1988qk}.
$\ket{B_\pm}$ states describe the ordered states breaking $\mathbb{Z}_2$ symmetry whereas $\ket{B_f}$ describes the $\mathbb{Z}_2$ symmetric gapped state as is becomes clear in the lattice counterpart. 

In open channel, the partition Euclidean path integral gives the partition function and in Ising CFT they are 
\begin{align}
&Z_{++}^{\text{open}} = Z_{--}^{\text{open}} = \chi_0, \\ \notag 
&Z_{+-}^{\text{open}} = Z_{-+}^{\text{open}} = \chi_{\frac{1}{2}} ,\notag \\
&Z_{ff}^{\text{open}} = \chi_0 + \chi_{\frac{1}{2}} \notag \\
&Z_{\pm f }^{\text{open}} = Z_{f\pm}=  \chi_{\frac{1}{16}}
\end{align}

We now turn to the lattice realization.
In the transverse field Ising model with boundaries, we consider the Hamiltonian
\begin{align}
H^{\text{open}}_{ab} = - \sum_{i = 1}^{L-1}Z_i Z_{i+1} -\sum_{i = 1}^L X_i - h_a Z_1 - h_b Z_L. 
\end{align}
For $a = \pm$, we take $h_\pm \to  \pm\infty$
This is the fixed boundary condition by the boundary magnetic field.
On the other hand, for $a =f$ we take $h_f = 0$ and we have free boundary conditions.

Lattice counterparts of the Cardy states are given by
\begin{align}
\ket{b_+} &= \otimes_{i = 1}^L\ket{+}_i \notag \\
\ket{b_-} &= \otimes_{i = 1}^L\ket{-}_i \notag \\ 
\ket{b_f} &= \otimes_{i = 1}^L\Big(\frac{\ket{+}_i+\ket{-}_i}{\sqrt{2}} \Big) \label{eq:TFIBS}
\end{align}
Those states are obtained by the infinitely massive limit of the critical Hamiltonian.
A massive deformation of the critical Ising model is 
\begin{align}
H = -  J \sum_{i=1}^L Z_i Z_{i+1} - g \sum_{i = 1}^L X_i.
\end{align}
These product states arise as infinitely massive limits of the ordered and disordered phases of the transverse-field Ising chain. 
In the limit $g/J\to0$, the two symmetry-broken ground states approach the fully polarized $Z$-basis product states $\ket{b_+}, \ket{b_-}$. 
In the opposite limit $J/g\to0$, the ground state approaches the fully polarized $X$-basis product state $\ket{b_f}$.
The first two states $\ket{b_+}, \ket{b_-}$ are ground states of $J\to \infty$ limit.

\section{Closed-string spectrum}
\begin{figure}[tb]
    \centering
    \includegraphics[width=\linewidth]{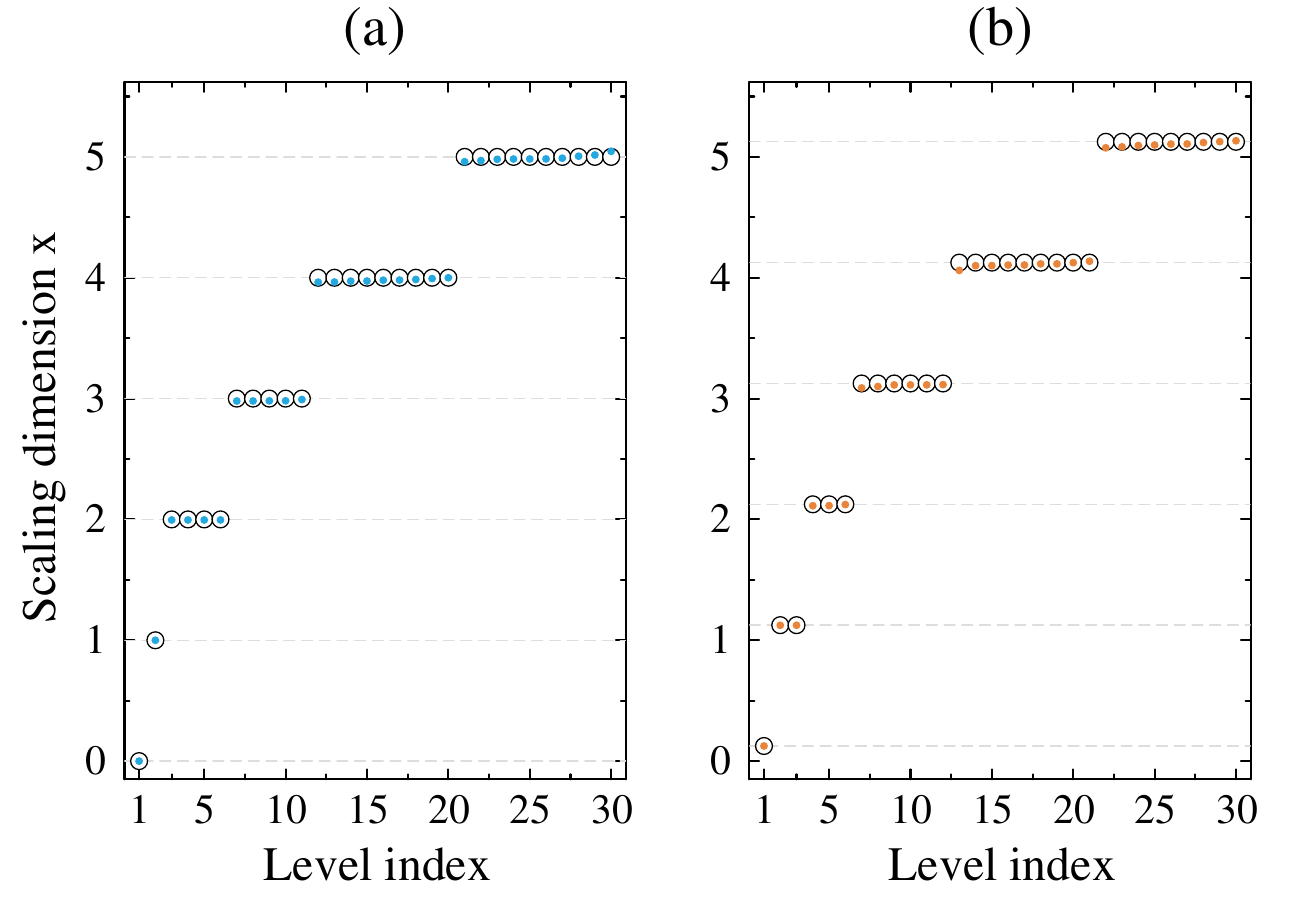}
    \caption{ The closed-string Ising CFT spectrum obtained from the transfer matrix spectrum of the thermal density operator. (a) and (b) represent the spectrum of the even and odd sectors. The colored dots and empty circles represent the numerical results and theoretical values, respectively.}
    \label{fig:closed_string}
\end{figure}

The temporal spectrum provides an equally powerful tool for extracting CFT data under periodic boundary conditions. The algorithm requires no essential modification: instead of evolving a state, we evolve a vectorized identity operator. This is achieved by preparing an MPS in a doubled local Hilbert space, $\mathbb{C}^d\otimes\tilde{\mathbb{C}}^d$, and applying imaginary-time evolution only to the physical space $\mathbb{C}^d$. After evolving for an imaginary time $\tau/2$, the transfer matrix of the resulting MPS encodes the thermal partition function $Z(\tau)=\mathrm{Tr}(e^{-\tau H})$. Figure~\ref{fig:closed_string} shows the numerically extracted spectrum for $\chi=80$ and $\tau=12$. The charge-resolved spectrum agrees with the CFT predictions with high precision. This provides a fast and accurate route to computing the excitation spectrum on a ring, in contrast to conventional MPS approaches, where accessing excited states typically requires more elaborate algorithms.

\section{Rotation of spacetime.}

\begin{figure}[tb]
    \centering
    \includegraphics[width=\linewidth]{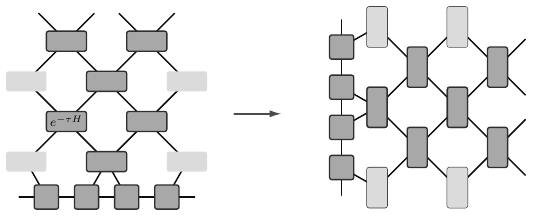}
    \caption{Spacetime duality on the lattice. (Left) Trotter decomposition of the imaginary-time evolution generated by a nearest-neighbor Hamiltonian yields a checkerboard tensor-network representation of the partition function. (Right) After a 90° rotation, the four-leg gates define the dual Hamiltonian, while the MPSs representing the initial and final states become boundary conditions.}
    \label{fig:90_rotation}
\end{figure}
The duality between the Hamiltonians under a $90^\circ$ spacetime rotation becomes transparent in the Trotterized path-integral representation, as illustrated in Fig.~\ref{fig:90_rotation}. We first discretize imaginary time using the Suzuki–Trotter decomposition, as shown in the left panel. For a Hamiltonian with nearest-neighbor interactions, such as the transverse-field Ising model, the resulting tensor network has a checkerboard structure. We then rotate the entire network by $90^\circ$, exchanging the spatial and imaginary-time directions. The rotated local tensors can now be interpreted as imaginary-time evolution gates of a dual Hamiltonian. Meanwhile, the MPSs representing the initial and final states are mapped to boundary interactions in the dual description, as shown in the right panel.

Let us elaborate on this concept on the Ising example: the Ising Hamiltonian consists of nearest-neighbor $ZZ$ terms and onsite $X$ terms. In a translationally invariant way, the local Trotter gates are generated by $$h = -JZ_iZ_{i+1} - \frac{g}{2}(X_i+X_{i+1}),$$ 
such that 
\begin{align}
e^{-\tau h}
=e^{\frac{\tau J}{2}Z_iZ_{i+1}}
e^{\frac{\tau g}{2} X_i}
e^{\frac{\tau g}{2} X_{i+1}}
e^{\frac{\tau J}{2}Z_{i}Z_{i+1}}
+O(\tau^3).
\end{align}
 Crucially, for the Ising model, $h_x=-JZZ$ is diagonal in the $Z$ basis. The $ZZ$ and $X$ terms can therefore be expressed as classical Boltzmann weights along the spatial and imaginary-time directions, respectively. Schematically, the symmetric Suzuki–Trotter decomposition takes the form
\begin{equation*}
    \includegraphics[width=0.9\linewidth]{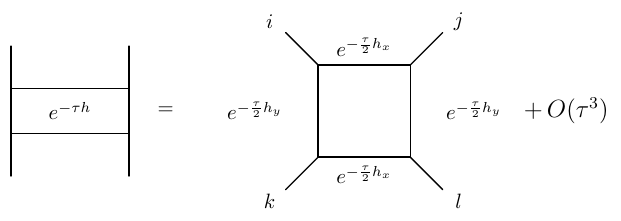},
\end{equation*}
where $h_x=-JZZ$ and $h_y=-gX$. The local transfer matrix in the imaginary-time ($y$) direction is given by
\begin{equation}
e^{\frac{g\tau}{2}X}
= \cosh\left(\frac{g\tau}{2}\right)I
+ \sinh\left(\frac{g\tau}{2}\right)X.
\end{equation}
Up to an overall normalization, this takes precisely the form of a classical Ising Boltzmann weight,
\begin{equation}
\begin{pmatrix}
e^{\tilde{K}/2} & e^{-\tilde{K}/2} \\
e^{-\tilde{K}/2} & e^{\tilde{K}/2}
\end{pmatrix},
\end{equation}
with the identification
\begin{equation}
e^{-\tilde{K}} = \tanh\left(\frac{g\tau}{2}\right).
\end{equation}
Thus, a $90^\circ$ rotation exchanges the spatial and imaginary-time directions, yielding another Ising transfer matrix with transformed couplings. This is closely related to Kramers–Wannier duality, under which the critical Ising model is self-dual. Furthermore, contracting the rotated local tensors with the boundary MPS tensors allows us to explicitly determine the boundary conditions in the dual description.

For more general models, however, the $90^\circ$ rotation should be understood as a rotation of the local four-leg tensors rather than a simple exchange of classical Boltzmann bonds. For example, the tensor network corresponding to the spin-$1/2$ XXZ chain is represented by six-vertex-model-type tensors, whose rotation provides the corresponding spacetime-dual description.

\section{Fidkowski--Kitaev  and Maldacena--Ludwig}
Here we summarize the Fidkowski--Kitaev Hamiltonian and Maldacena--Ludwig boundary state.

The Kitaev chain Hamiltonian is 
\begin{align}
H_{8 \text{Kitaev}} = \frac{i}{2} \sum_{\ell=1}^L \sum_{a=1 }^8 \chi_{\ell}^a \chi_{\ell+1}^a.
\end{align}
The lattice counterpart of the free boundary states $\ket{\Phi_{free}}$  are given by the ground states of the Hamiltonian 
\begin{align} 
H=i \sum_{\ell=1}^L \sum_{a=1 }^8 \frac{1 + m(-1)^{\ell}}{2} \chi_{\ell}^a \chi_{\ell+1}^a, \quad m = \pm 1 .
\end{align}
For this Hamiltonian, either even $\ell$ or odd $\ell$ terms vanish depending on the sign, and the ground state is completely dimerized.
After the Jordan--Wigner transformation, each layer $a$ is mapped to the Ising model, and those dimerized states are the fermionic counterparts of the lattice Ising boundary states \eqref{eq:TFIBS} \cite{Ebisu:2021acm}.

The Fidkowski--Kitaev Hamiltonian is defined for eight Majorana fermion system $\{ \chi^a ,\chi^b\} = 2\delta^{ab}$ by
\begin{align}
H _{FK} &= -\sum_{a<b<c<d} \Phi_{abcd} \chi^a\chi^b\chi^c \chi^d, \notag \\
\Phi_{abcd}&=\braket{\psi_{FK}|\chi^{[a}\chi^b\chi^c \chi^{d]}|\psi_{FK}},
\end{align}
Here $\Phi_{abcd}$ is the so-called Cayley 4-form \cite{Shatashvili:1994zw}.
The Cayley $4-$ form is invariant under the $\mathrm{spin}(7) \subset SO(8)$ subgroup which stabilizes the Fidkowski--Kitaev state.

The Cayley 4-form also appears in the Maldacena--Ludwig boundary state through the triality automorphisms.
Using the Cayley 4-form, the triality automorphism  $\omega$ is
\begin{align}
\omega(J^{ab}) = \frac{1}{2}J^{ab} + \frac{1}{4} \sum_{cd} \Phi_{abcd}J^{cd} .
\end{align}
This action satisfies $\omega(\omega(J)) = J$ and eigenvalue $1$ part $\omega (J^{ab}) = J^{ab}$ gives the $Spin(7)$ subgroup.

The branching rules under the conformal embedding
$\mathfrak{so}(8)_1 \supset
\mathfrak{so}(7)_1 \times \mathrm{Ising}$ are
\begin{align}
\bm{0}   &\to (\bm{0},I)\oplus(\bm{7},\epsilon), \notag\\
\bm{8}_v &\to (\bm{8},\sigma), \notag\\
\bm{8}_s &\to (\bm{8},\sigma), \notag\\
\bm{8}_c &\to (\bm{7},I)\oplus(\bm{0},\epsilon).
\end{align}
Here each pair labels a representation of
$\mathfrak{so}(7)_1$ and an Ising primary respectively,
where $\bm{7}$ is the vector representation of $\mathrm{Spin}(7)$ and $\bm{8}$ is the spinor representation.
Then, the free boundary states are 
\begin{align}
\ket{B_{\pm}}_{NS} = \ket{0}\rangle \pm \ket{\bm{8}_v}\rangle  = \ket{\bm{0},I}\rangle + \ket{\bm{7},\epsilon}\rangle \pm \ket{\bm{8},\sigma}\rangle 
\end{align}
Noticing that $\omega$ acts as the flip of $\epsilon \to -\epsilon$  in the Ising field part, the Maldacena--Ludwig boundary state is 
\begin{align}
\ket{ML}_{\mathrm{NS}} = \sqrt{2}\ket{0}\rangle_{\omega} = \sqrt{2} (\ket{\bm{0},I}\rangle-\ket{\bm{7},\epsilon}\rangle  ),
\end{align}
Then, we obtain the closed-channel amplitudes 
\begin{align}
&_{\mathrm{NS}}\braket{ML|\tilde{q}^{L_0+ \tilde{L}_0 - \frac{c}{24}}| ML}_{\mathrm{NS}} \notag \\
&= 2(\chi_0^{SO(7)_1}(\tilde{q})\chi_0^{\text{Ising}}(\tilde{q}) + \chi_{\bm{7}}^{SO(7)_1}(\tilde{q})\chi_\epsilon^{\text{Ising}}(\tilde{q}))  \notag \\
&= 2\chi_0^{SO(8)_1}(\tilde{q})
\end{align}
and 
\begin{align}
&_{\mathrm{NS}}\braket{ML|\tilde{q}^{L_0+ \tilde{L}_0 - \frac{c}{24}}| B_{\pm}}_{\mathrm{NS}} \notag \\
&= \sqrt{2}(\chi_0^{SO(7)_1}(\tilde{q})\chi_0^{\text{Ising}}(\tilde{q}) - \chi_{\bm{7}}^{SO(7)_1}(\tilde{q})\chi_\epsilon^{\text{Ising}}(\tilde{q})) .
\end{align}
Applying the modular transformation to go to the open channel, we obtain the partition function $Z_{ML,ML}$
 and $Z_{ML,free}$.
 The action of chiral  fermion parity $(-1)^{F_L}$ on the Ishibashi state is 
 \begin{align}
 (-1)^{F_L}\ket{0,I}\rangle  &= \ket{0,I}\rangle \notag \\
 (-1)^{F_L}\ket{\bm{7},\epsilon}\rangle  &= \ket{\bm{7},\epsilon}\rangle \notag \\
 (-1)^{F_L}\ket{\bm{8},\sigma}\rangle  &= -\ket{\bm{8},\sigma}\rangle .
 \end{align}
 From this action, we can clearly see that the Maldacena--Ludwig boundary state is invariant under the $(-1)^{F_L}$ operation, whereas two free boundary states break the chiral fermion parity $(-1)^{F_L}$  and are exchanged by it.

 \section{Conformal weight data for Haagerup}
The Hamiltonian of the $\mathcal{H}_3$ Haagerup critical model is 
We tabulate the data corresponding to Fig.~\ref{fig:haagerup_spectrum} in Tables \ref{tab:gridAA} and \ref{tab:gridAB}.
We used a fixed multiplicity $\chi_{s} = 160$ which represents the number of states kept in each symmetry sector.
We denote in bold the primaries, which we exclude to be Virasoro descendants of a lower level within $\pm 0.05$.
The $\dagger$ notation means that this value changes by more than $\pm 0.02$ from the $\chi_s = 120$ simulation, which is the typical case for higher levels and may reflect a poorer convergence.
Potential primaries with $\dagger$ are not in bold, as we cannot conclusively state this.

\begin{table}[tp]
\caption{A--A boundary weights $h$ per symmetry sector. Charge $\alpha^2$ equals $\alpha$ exactly.}\label{tab:gridAA}
\begin{ruledtabular}
\begin{tabular}{rlllll}
$n$ & $1$ & $\alpha,\alpha^2$ & $\rho$ & $\alpha\rho$ & $\alpha^2\rho$\\
\hline
1 & $\mathbf{0.000}\phantom{^\dagger}$ & $\mathbf{2.405}\phantom{^\dagger}$ & $\mathbf{0.683}\phantom{^\dagger}$ & $\mathbf{0.648}\phantom{^\dagger}$ & $\mathbf{1.108}\phantom{^\dagger}$\\
2 & $2.000\phantom{^\dagger}$ & $\mathbf{3.164}\phantom{^\dagger}$ & $1.681\phantom{^\dagger}$ & $1.651\phantom{^\dagger}$ & $\mathbf{1.677}\phantom{^\dagger}$\\
3 & $\mathbf{2.724}\phantom{^\dagger}$ & $\mathbf{3.341}\phantom{^\dagger}$ & $\mathbf{1.773}\phantom{^\dagger}$ & $\mathbf{1.845}\phantom{^\dagger}$ & $2.091\phantom{^\dagger}$\\
4 & $3.010\phantom{^\dagger}$ & $3.409\phantom{^\dagger}$ & $\mathbf{2.071}\phantom{^\dagger}$ & $2.630\phantom{^\dagger}$ & $\mathbf{2.389}\phantom{^\dagger}$\\
5 & $\mathbf{3.154}\phantom{^\dagger}$ & $\mathbf{4.046}\phantom{^\dagger}$ & $\mathbf{2.305}\phantom{^\dagger}$ & $2.652\phantom{^\dagger}$ & $2.681\phantom{^\dagger}$\\
6 & $3.751^\dagger$ & $4.197^\dagger$ & $\mathbf{2.637}\phantom{^\dagger}$ & $\mathbf{2.733}\phantom{^\dagger}$ & $\mathbf{2.681}\phantom{^\dagger}$\\
7 & $\mathbf{3.828}\phantom{^\dagger}$ & $4.377^\dagger$ & $2.679\phantom{^\dagger}$ & $2.839\phantom{^\dagger}$ & $\mathbf{2.902}\phantom{^\dagger}$\\
8 & $4.030\phantom{^\dagger}$ & $4.394^\dagger$ & $2.681\phantom{^\dagger}$ & $\mathbf{2.856}\phantom{^\dagger}$ & $\mathbf{3.057}\phantom{^\dagger}$\\
9 & $4.080^\dagger$ & $4.501^\dagger$ & $\mathbf{2.732}\phantom{^\dagger}$ & $\mathbf{3.003}\phantom{^\dagger}$ & $3.096^\dagger$\\
10 & $4.206^\dagger$ & $4.769^\dagger$ & $2.777\phantom{^\dagger}$ & $\mathbf{3.151}\phantom{^\dagger}$ & $3.398\phantom{^\dagger}$\\
11 & $4.218^\dagger$ & $4.857^\dagger$ & $3.084\phantom{^\dagger}$ & $3.608\phantom{^\dagger}$ & $\mathbf{3.514}\phantom{^\dagger}$\\
12 & $4.756^\dagger$ & $5.146^\dagger$ & $3.321^\dagger$ & $3.642^\dagger$ & $\mathbf{3.577}\phantom{^\dagger}$\\
\end{tabular}
\end{ruledtabular}
\end{table}

\begin{table}[tp]
\caption{A--TA boundary weights $h$ per symmetry sector. Charge $\alpha^2$ equals $\alpha$ to $2\times10^{-3}$.}\label{tab:gridAB}
\begin{ruledtabular}
\begin{tabular}{rlllll}
$n$ & $1$ & $\alpha,\alpha^2$ & $\rho$ & $\alpha\rho$ & $\alpha^2\rho$\\
\hline
1 & $\mathbf{0.912}\phantom{^\dagger}$ & $\mathbf{2.050}\phantom{^\dagger}$ & $\mathbf{0.238}\phantom{^\dagger}$ & $\mathbf{1.239}\phantom{^\dagger}$ & $\mathbf{0.990}\phantom{^\dagger}$\\
2 & $1.915\phantom{^\dagger}$ & $\mathbf{2.984}\phantom{^\dagger}$ & $1.235\phantom{^\dagger}$ & $\mathbf{1.628}\phantom{^\dagger}$ & $\mathbf{1.941}\phantom{^\dagger}$\\
3 & $2.921\phantom{^\dagger}$ & $3.043\phantom{^\dagger}$ & $\mathbf{1.589}\phantom{^\dagger}$ & $2.226\phantom{^\dagger}$ & $1.993\phantom{^\dagger}$\\
4 & $2.933\phantom{^\dagger}$ & $\mathbf{3.945}\phantom{^\dagger}$ & $2.204\phantom{^\dagger}$ & $\mathbf{2.343}\phantom{^\dagger}$ & $\mathbf{2.223}\phantom{^\dagger}$\\
5 & $\mathbf{3.057}\phantom{^\dagger}$ & $4.013^\dagger$ & $2.232\phantom{^\dagger}$ & $2.633\phantom{^\dagger}$ & $\mathbf{2.814}\phantom{^\dagger}$\\
6 & $\mathbf{3.835}\phantom{^\dagger}$ & $4.054\phantom{^\dagger}$ & $\mathbf{2.403}\phantom{^\dagger}$ & $\mathbf{2.720}\phantom{^\dagger}$ & $2.933\phantom{^\dagger}$\\
7 & $\mathbf{3.852}\phantom{^\dagger}$ & $4.081^\dagger$ & $2.594\phantom{^\dagger}$ & $\mathbf{2.812}\phantom{^\dagger}$ & $\mathbf{2.966}\phantom{^\dagger}$\\
8 & $3.956^\dagger$ & $\mathbf{4.181}\phantom{^\dagger}$ & $\mathbf{2.754}\phantom{^\dagger}$ & $3.200\phantom{^\dagger}$ & $2.994\phantom{^\dagger}$\\
9 & $4.015^\dagger$ & $4.575^\dagger$ & $\mathbf{3.037}\phantom{^\dagger}$ & $3.236^\dagger$ & $3.006^\dagger$\\
10 & $4.099^\dagger$ & $4.981^\dagger$ & $3.209\phantom{^\dagger}$ & $3.342\phantom{^\dagger}$ & $3.217\phantom{^\dagger}$\\
11 & $\mathbf{4.285}\phantom{^\dagger}$ & $5.028^\dagger$ & $3.243\phantom{^\dagger}$ & $\mathbf{3.424}\phantom{^\dagger}$ & $\mathbf{3.452}\phantom{^\dagger}$\\
12 & $4.895^\dagger$ & $5.058^\dagger$ & $3.248\phantom{^\dagger}$ & $\mathbf{3.574}\phantom{^\dagger}$ & $\mathbf{3.537}\phantom{^\dagger}$\\
\end{tabular}
\end{ruledtabular}
\end{table}

\end{document}